\documentclass[]{spie}  %>>> use for US letter paper
\usepackage{amsmath,amsfonts,amssymb}
\usepackage{graphicx}
\usepackage[colorlinks=true, allcolors=blue]{hyperref}

\title{Towards understanding interactions between the AO system and segment co-phasing with the vector-Zernike wavefront sensor on Keck}

\author[a]{Ma\"issa Salama}
\author[b]{Charlotte Guthery}
\author[a]{Vincent Chambouleyron}
\author[a]{Rebecca Jensen-Clem}
\author[c]{J. Kent Wallace}
\author[c]{Mitchell Troy}
\author[b]{Jacques-Robert Delorme}
\author[d]{Daren Dillon}
\author[e]{Daniel Echeverri}
\author[e]{Yeyuan (Yinzi) Xin}
\author[e]{Wen Hao (Jerry) Xuan}
\author[e]{Nemanja Jovanovic}
\author[e]{Dimitri Mawet}
\author[b]{Peter L. Wizinowich}
\author[a]{Rachel Bowens-Rubin}

\affil[a]{University of California, Santa Cruz, California, U.S.A.}
\affil[b]{W. M. Keck Observatory, Hawaii, U.S.A.}
\affil[c]{Jet Propulsion Laboratory, California Institute of Technology, U.S.A.}
\affil[d]{University of California Observatory, California, U.S.A.}
\affil[e]{California Institute for Technology, California, U.S.A.}

\authorinfo{Further author information: (Send correspondence to M.S.)\\M.S.: E-mail: msalama@ucsc.edu}

\begin{document} 
\maketitle

\begin{abstract}
%We demonstrated the first on-sky primary mirror segment closed-loop control on Keck using a vector-Zernike wavefront sensor (vZWFS) and improved the Strehl ratio on the NIRC2 science camera by up to 10 percentage points. Segment co-phasing errors are a primary contributor to Keck contrast limits and will be necessary to correct for the segmented ELTs. The goal of the post-AO ZWFS on Keck is to monitor and correct segment co-phasing errors in parallel with science observations. The ZWFS is ideal for measuring phase discontinuities and is the most sensitive WFS, but has limited dynamic range. The Keck vZWFS consists of a metasurface mask imposing two different phase shifts to orthogonal polarizations, extending its dynamic range. We report on the vZWFS closed-loop co-phasing performance and facilitization at Keck. This work is supported by the vZWFS on the SEAL testbed at UCSC. We discuss the results in the context of high-contrast imaging.

We extend our previous demonstration of the first on-sky primary mirror segment closed-loop control on Keck using a vector-Zernike wavefront sensor (vZWFS), which improved the Strehl ratio on the NIRC2 science camera by up to 10 percentage points. Segment co-phasing errors contribute to Keck contrast limits and will be necessary to correct for the segmented Extremely Large Telescopes and future space missions. The goal of the post-AO vZWFS on Keck is to monitor and correct segment co-phasing errors in parallel with science observations. The ZWFS is ideal for measuring phase discontinuities and is one of the most sensitive WFSs, but has limited dynamic range. The Keck vZWFS consists of a metasurface mask imposing two different phase shifts to orthogonal polarizations, split into two pupil images, extending its dynamic range. We report on the vZWFS closed-loop co-phasing performance and early work towards understanding the interactions between the AO system and segment phasing. We discuss a comparison of the AO performance when co-phasing by aligning segment edges, as is currently done at Keck, compared with aligning to the average phase over the segments, as is done by the vZWFS.

\end{abstract}


\section{Introduction}
\label{sec:intro} 

As we approach the era of next generation Extremely Large Telescopes (ELTs) and space missions, with the goal of directly imaging an Earth-like planet, fine phase control is increasingly important. Segment co-phasing errors in segmented apertures, which is the necessary design for the next generation of telescopes, both on the ground and in space, are therefore crucial to understand and correct. The W. M. Keck Observatory is currently the only ground-based segmented telescope with an adaptive optics (AO) system and high-contrast imaging science instruments. It is therefore ideal for technology development and testing methods relevant for the fine phase control and segment co-phasing that will be required for ELTs and future space missions, such as the Habitable Worlds Observatory (HWO).

The Zernike wavefront sensor (ZWFS), one of the most photon-efficient wavefront sensors, is also sensitive to discontinuous phase aberrations, such as those caused by a segmented aperture. At the Keck II telescope, a ZWFS is installed\cite{Salama24} in the Keck Planet Imager and Characterizer (KPIC)\cite{Mawet16,Delorme21} instrument, downstream from the AO system\cite{Wizinowich13}. However, one of the main limitations of the ZWFS is its small dynamic range. A vector-Zernike mask increases the dynamic range by simultaneously applying two different phase shifts to orthogonally polarized light\cite{Doelman19}, which is then split into two pupil images. The ZWFS at Keck uses a vector-Zernike mask made of metasurface material\cite{Wallace23}. In Salama et al. (2024)\cite{Salama24}, we reported on the first on-sky closed-loop control of the Keck primary mirror segments using this vector-ZWFS (vZWFS) which resulted in an increase in the Strehl ratio (SR), by up to ten percentage points, on the NIRC2 science camera. 

In this work, building on the first results presented in Salama et al. (2024)\cite{Salama24}, we present additional tests conducted with the vZWFS on Keck, and the impact on post-AO image quality. In section \S\ref{sec:PrimCL}, we present results from four new closed-loop tests on the primary mirror segment pistons. In section \S\ref{sec:PCS} we compare the vZWFS segment co-phasing result with the segment co-phasing method usually performed at Keck using edge sensors and the phasing camera system (PCS). In addition, we begin exploring how to disentangle sources of aberrations between the primary mirror co-phasing errors and other AO residuals and systematics by using the vZWFS signal to send slope-offsets to the Shack Hartmann WFS in the main AO loop (section \S\ref{sec:K2AO}). Finally, we discuss other diagnostic possibilities with the vZWFS such as searching for vibrations in the system by taking high-speed datasets with the vZWFS (section \S\ref{sec:Fast}). We conclude in section \S\ref{sec:Concl} with a discussion of next steps towards determining the optimal methods of segment co-phasing and fine phase control for increased AO performance. With the benefit of simultaneously improving contrast limits at Keck in the short-run, while working towards the long-term goal of testing and developing control architectures that utilize both post-AO diagnostics, such as the ZWFS, as well as edge sensor data for the next generation of ELTs and space missions.

\section{On-sky Primary mirror segment closed-loop results}
\label{sec:PrimCL} 

We report on four closed-loop tests conducted in 2024 February using the vZWFS on Keck II to control the primary mirror segment pistons. For these observations we ran daytime calibrations, which consist of the normal Keck AO bench calibrations (to correct for NCPAs between the SHWFS and NIRC2), then we corrected for NCPAs between the vZWFS arm in KPIC and NIRC2 by sending a set of static slope offsets to the SHWFS after closing the loop on the vZWFS signal\footnote{Unlike in Salama et al. (2024)\cite{Salama24}, the DM inside of KPIC was not working during our February vZWFS runs, we therefore could not compensate for NCPAs using that DM. For this reason, NCPAs were now present on NIRC2. Instead, there is the option to switch between pupil and focal mode on the CRED2 detector in KPIC. We therefore alternated between the pupil mode for the vZWFS images and closed-loop runs, and focal mode to take before and after PSF images on the CRED2 camera.}. Once on-sky, with the AO loops closed, for each primary mirror closed-loop test, the process is as follows: 1) we take five vZWFS datasets, consisting of 60 frames taken at 2Hz and corresponding to 30 seconds integration time, 2) for each dataset, we average the 60 frames to remove non-static residuals, then reconstruct the phase using a nonlinear reconstructor based on a numerical model and iterative algorithm (see Chambouleyron et al. 2024\cite{Chambouleyron24} in this conference for a comparison of different phase reconstruction techniques for the vZWFS), 3) after removing global piston, tip, tilt, and focus, we then extract each segment piston value, 4) then average the extracted segment pistons across the five datasets, 5) send the average piston measurements for each segment as offsets to the Keck segments through the Active Control System (ACS)\cite{Cohen94}. We then repeat steps 1-5 for five to six iterations, which is when we stop seeing a decrease in the root mean square (RMS) of the measured segment piston values. 

\begin{table}[h!]
    \centering
    \begin{tabular}{c|c|c|c|c|c|c }
        Run & Date (UT) & Star & H-mag & Elevation range [deg] & Pupil Rotation [deg] & Seeing range \\
        \hline
        1 & 2024-02-21 & Denebola & 1.9 & $65 - 70$ & 0 & $0.43''-0.59''$ \\
        2 & 2024-02-22 & SAO117885 & 4.2 & $65 - 68$ & 0 & $0.34''-0.53''$ \\
        3 & 2024-02-22 & SAO117885 & 4.2 & $58 - 62$ & 0 & $0.41''-0.57''$ \\
        4 & 2024-02-22 & HIP 62808 & 4.5 & $63 - 68$ & 3.73 & $0.34''-0.62''$ \\
        \hline
    \end{tabular}
    \caption{Summary of observing conditions during the four closed loop tests. The seeing measurements are reported by the CFHT DIMM seeing monitor.}
    \label{tab:obs}
\end{table}

\begin{figure}
    \centering
    \includegraphics[height=0.7\textheight]{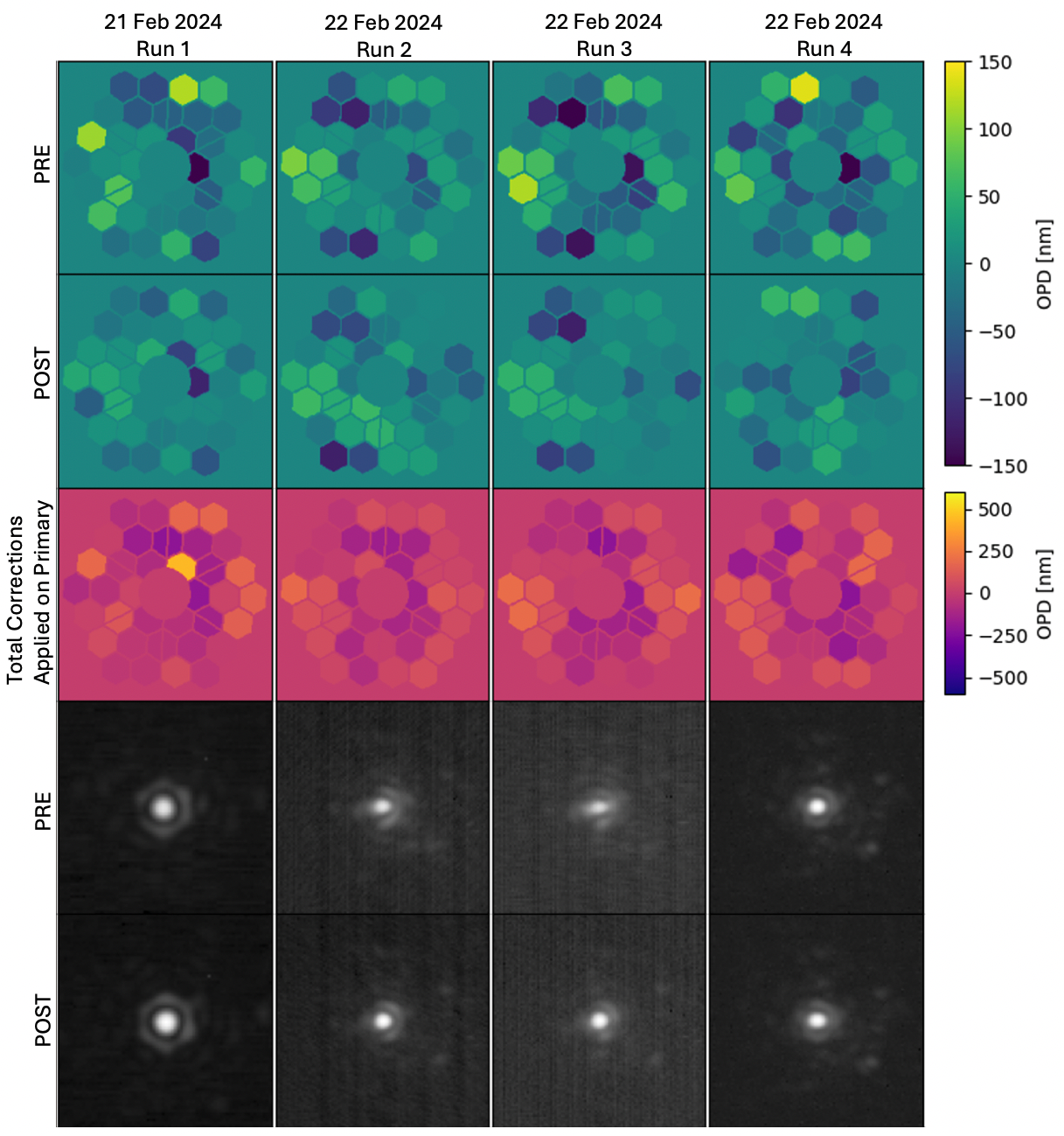}
    \caption{%The first four closed-loop tests are those reported in Salama et al. (2024) from 2023 August 2, 4, and 13. And the last four closed-loop tests are those from this work, conducted on 2024 February 21 and 22. 
    Summaries of the four closed-loop runs reported in this work are shown. The top two rows show the before and after segment piston values measured by the vZWFS. The middle row shows maps of the total piston corrections applied on the primary mirror throughout the closed-loop run. The bottom two rows show the corresponding before and after PSF images, displayed in square-root stretch, and were taken with the CRED2 camera in KPIC in the H-band. The corresponding SRs and RMSs are summarized in Table \ref{tab:delta_SRs}.}
    \label{fig:CL_summary}
\end{figure}

On 2024 February 21, we ran one closed-loop test, and three more closed-loop tests the next night on 2024 February 22. The observing details for the four closed-loop runs are summarized in Table \ref{tab:obs}. Figure \ref{fig:CL_summary} summarizes the four closed-loop test results, displaying the segment piston value maps (in wavefront OPD) measured by the vZWFS before and after the closed-loop run, as well as the map of the total piston corrections applied to the primary mirror throughout each closed-loop run. Finally, the corresponding before and after PSF images, taken in H-band on the CRED2 camera, are shown. Table \ref{tab:delta_SRs} summarizes the SRs and the RMSs of the total segment piston corrections applied on the primary mirror throughout each closed-loop run.

\begin{table}
    \centering
    \begin{tabular}{c|c|c|c|c|c}
        % &  &  &  &  &  & Total Corrections \\
        Run & Pre Closed-Loop SR & Post Closed-Loop SR & $\Delta$SR & $\Delta$RMS & Total Corrections Applied \\
        \hline
        1 & 43.4\% $\pm$ 4.7\% & 50.7\% $\pm$ 3.4\% & 7.3\% $\pm$ 5.8\% & 103 nm & 157 nm RMS \\
        2$^*$ & 23.3\% $\pm$ 2.3\% & 32.7\% $\pm$ 2.0\% & 9.4\% $\pm$ 3.0\% & 153 nm & 90 nm RMS \\
        3 & 18.6\% $\pm$ 4.0\% & 32.9\% $\pm$ 3.9\% & 14.3\% $\pm$ 5.6\% & 198 nm & 106 nm RMS \\
        4$^{*\dagger}$ & 41.8\% $\pm$ 8.5\% & 51.1\% $\pm$ 1.5\% & 9.3\% $\pm$ 8.6\% & 117 nm & 128 nm RMS \\ 
        %4 & 41.8\% $\pm$ 8.5\% & 50.8\% $\pm$ 1.1\% & 9.1\% $\pm$ 8.6\% & 116 nm & 147 nm RMS \\ 
        \hline
    \end{tabular}
    \caption{Strehl Ratio differences and corresponding changes in RMS wavefront (OPD) between CRED2 images taken before and after each closed-loop test. The last column shows the total RMS of the piston commands (in wavefront OPD) sent to the primary mirror segments by the end of each vZWFS closed-loop test. $^*$The starting shape on the primary mirror for these runs was the one just phased the same night. $^\dagger$The pupil rotation angle for this run was 3.73 degrees, as opposed to 0 degrees for the others. Details about the different set-ups for Runs 2 - 4 are explained in sections \S\ref{sec:PCS} and \S\ref{sec:K2AO}.}
    \label{tab:delta_SRs}
\end{table}

The average length of our eight closed-loop runs was 20 minutes, $\sim$4 minutes per iteration. This is currently a manual process and could be a lot faster once automated, with the limiting factor being the amount of desired time for integration over which to average the atmospheric residuals. The phase reconstruction algorithm and segment piston extraction takes seconds to run and applying the corrections to the primary mirror also take seconds to settle. If the closed-loop process were automated and if one 30-second image were taken per iteration (as opposed to our current process of taking five images and averaging the resulting piston measurements for each iteration), the length of the closed-loop run could be $\lesssim$5 minutes, though the average measurement uncertainty would likely increase from 11~nm RMS \cite{Salama24} to $\sim$24~nm RMS.

\section{Comparison of vector-Zernike wavefront sensing and edge sensing for primary mirror segment co-phasing}
\label{sec:PCS} 

The normal process for phasing the Keck segments uses PCS\cite{Chanan94} with edge sensors and thus co-phases the segments by aligning their edges. In contrast, the vZWFS co-phases the segments by measuring the average phase over each segment, after AO correction with the deformable mirror (DM) and minimizes the difference between these segment average phases by pistoning the segments. Since the segments are not flat, these two techniques are not equivalent and likely lead to different phasing solutions. In fact these two solutions are far from equivalent, the residual edge phasing errors from segment aberrations is $\lesssim$ 175 nm RMS OPD. We are interested in determining which method yields the best AO performance and thus highest contrasts for post-AO coronagraphic observations. We conducted a first test to begin investigating this question on 2024 February 22. The telescope had last been phased on 2024 January 28 (which we will refer to as ``PCS0128"). On February 22, we ran PCS to re-phase the telescope (``PCS0222"). The RMS of the difference in piston values between PCS0128 and PCS0222 is 85 nm OPD, as measured by PCS. When measuring the piston values from PCS0222 at the beginning and end of night, the RMS of the difference in pistons was 55 nm RMS, and the noise of the difference of two measurements is typically 85 nm RMS. Therefore, the mirror phasing was stable throughout the night.

In order to compare with the vZWFS-phasing, we ran two closed-loop tests, controlling the primary mirror pistons with the vZWFS. One closed-loop run started from the nominal telescope phasing, PCS0128, and yielded the ZWFS-phasing solution ``ZWFS3" (Run 3 in Tables \ref{tab:obs} \& \ref{tab:delta_SRs}). For the other closed-loop run, we started from the just-phased telescope by PCS (PCS0222), and yielded the corresponding vZWFS solution ``ZWFS2" (Run 2 in Tables \ref{tab:obs} \& \ref{tab:delta_SRs}). In both cases, the vZWFS increased the PSF SR on the CRED2 camera, suggesting that fine-tuning the phasing with the vZWFS improves the AO performance. Furthermore, when measuring with PCS the two phasing solutions generated by the vZWFS with the two PCS phasing starting points, we find that 1) the two phasing runs yielded the same phasing solution, with the RMS of the difference of the piston values between ZWFS3 and ZWFS2 being 55 nm, and 2) the vZWFS phasing solution is different from the PCS phasing solution, with the RMS of the difference of piston values between ZWFS2 and PCS0222 being 85 nm. These results are summarized in Table \ref{tab:PCS}. We hope to re-execute these tests using a more precise phasing method in PCS that has an RMS piston error of 12 vs 60 nm (OPD).

\begin{table}[h]
    \centering
    \begin{tabular}{c|c}
        Phasing Solution Comparisons & RMS of Segment Piston Differences [nm, OPD]\\
        \hline
        PCS0222 - PCS0128 & 85 \\
        ZWFS3 - PCS0128 & 140 \\
        ZWFS2 - PCS0222 & 85 \\
        ZWFS2 - ZWFS3 & 55 \\
        PCS0222 - PCS0222 & 55 (same solution, measured at beginning and end of night) \\
        \hline
    \end{tabular}
    \caption{Comparing the segment piston values, as measured by PCS, for different phasing solutions. The two vZWFS closed-loop runs yielded similar solutions and the vZWFS solution differed slightly from the PCS solution.}
    \label{tab:PCS}
\end{table}

\section{Towards disentangling primary mirror co-phasing errors from other AO effects}
\label{sec:K2AO} 

In order to begin identifying the sources of aberrations measured by the ZWFS, separating those originating at the primary mirror from those resulting from AO residuals and systematics, we tested closing the loop on the Xinetics DM in the Keck AO loop, by sending slope offsets to the SHWFS. The objective is to test correcting aberrations that represent global modes across the full pupil with the Xinetics DM and correcting mid-spatial frequency modes from segment co-phasing errors with the primary mirror in order to compare the resulting PSF quality and determine the best strategy for maximizing the SRs and achievable post-coronagraphic contrasts.

\begin{figure}
    \centering
    \includegraphics[width=0.7\textwidth]{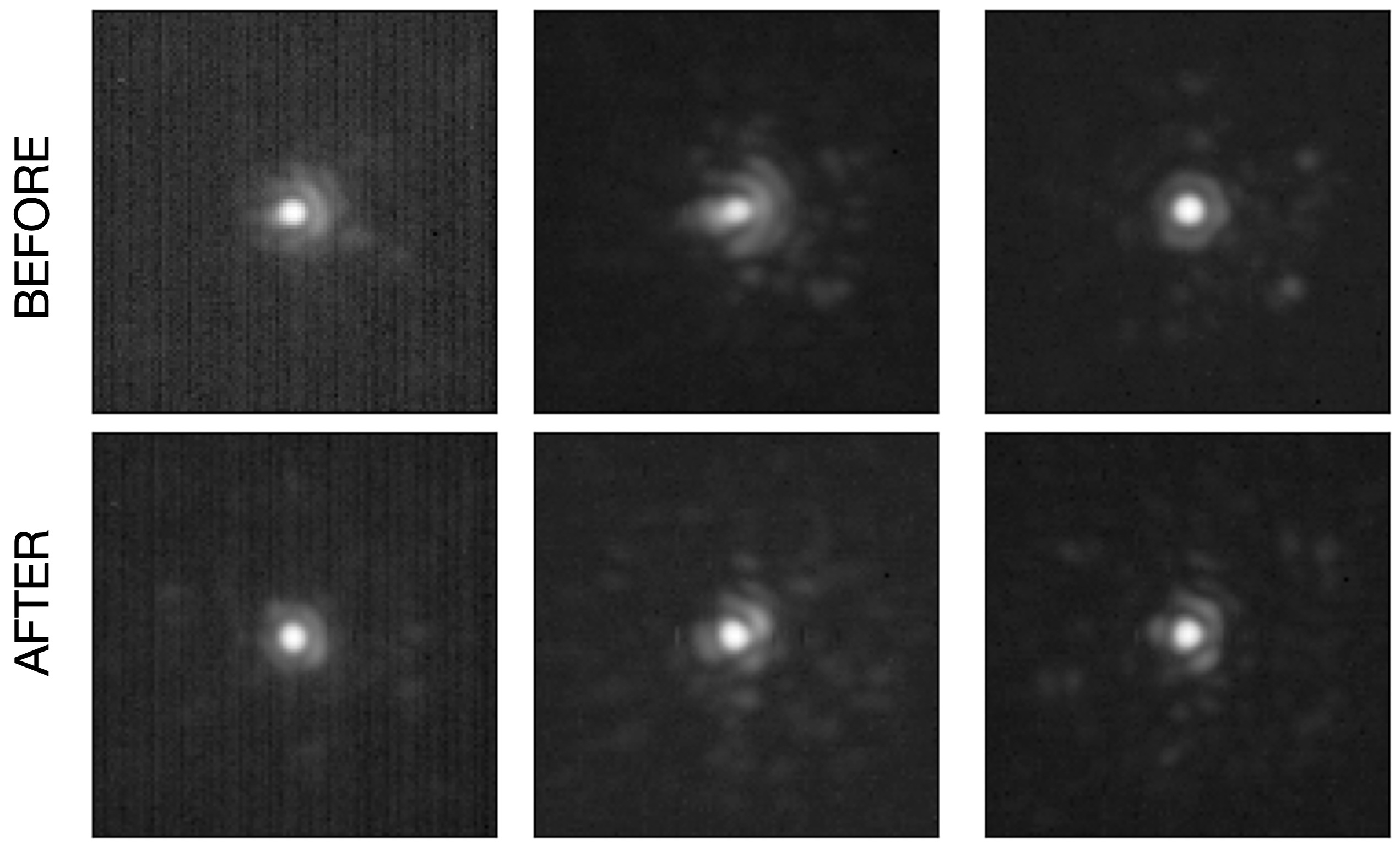}
    \caption{PSF images displayed in square-root stretch taken in the H-band on the CRED2 detector in KPIC. Top row are the images corresponding to before the sending slope offsets to the SHWFS in closed-loop from the vZWFS signal, and bottom row are after.}
    \label{fig:Xinetics_CL}
\end{figure}

We ran three closed-loop tests controlling the Xinetics DM, one on 21 February 2024 and two the next night on 22 February 2024. The before and after PSF images taken on the CRED2 camera in KPIC are shown in Figure \ref{fig:Xinetics_CL}. The first two runs resulted in an increase in the PSF SR, and the third run did not. However, that third run started with an already good PSF and was done following the vZWFS closed-loop on the primary mirror (after Run 4), highlighting the importance of developing control architectures for determining the correct sequence and signal interpretation when combining control of both the primary mirror segment pistons and sending offsets in closed-loop to the AO system.

Two of the eight primary mirror segment closed-loop runs were done with the pupil rotation at 3.73 degrees (which is the angle required for observing with the vortex coronagraph on NIRC2), while the other runs were done with the pupil rotation at 0 degrees. In the four closed-loop runs presented in this work, Run 4 was done with the pupil rotation at 3.73 degrees. Further tests will be needed to investigate the effect of pupil rotation on the interaction between the AO correction loop and segment phasing. It is likely that if certain SHWFS subapertures line up across two segment edges, and if those have differential pistons, that the SHWFS measures a local tip/tilt and may already be correcting it with the Xinetics DM.

\section{High speed}
\label{sec:Fast} 

\begin{figure}
    \centering
    \includegraphics[width=\textwidth]{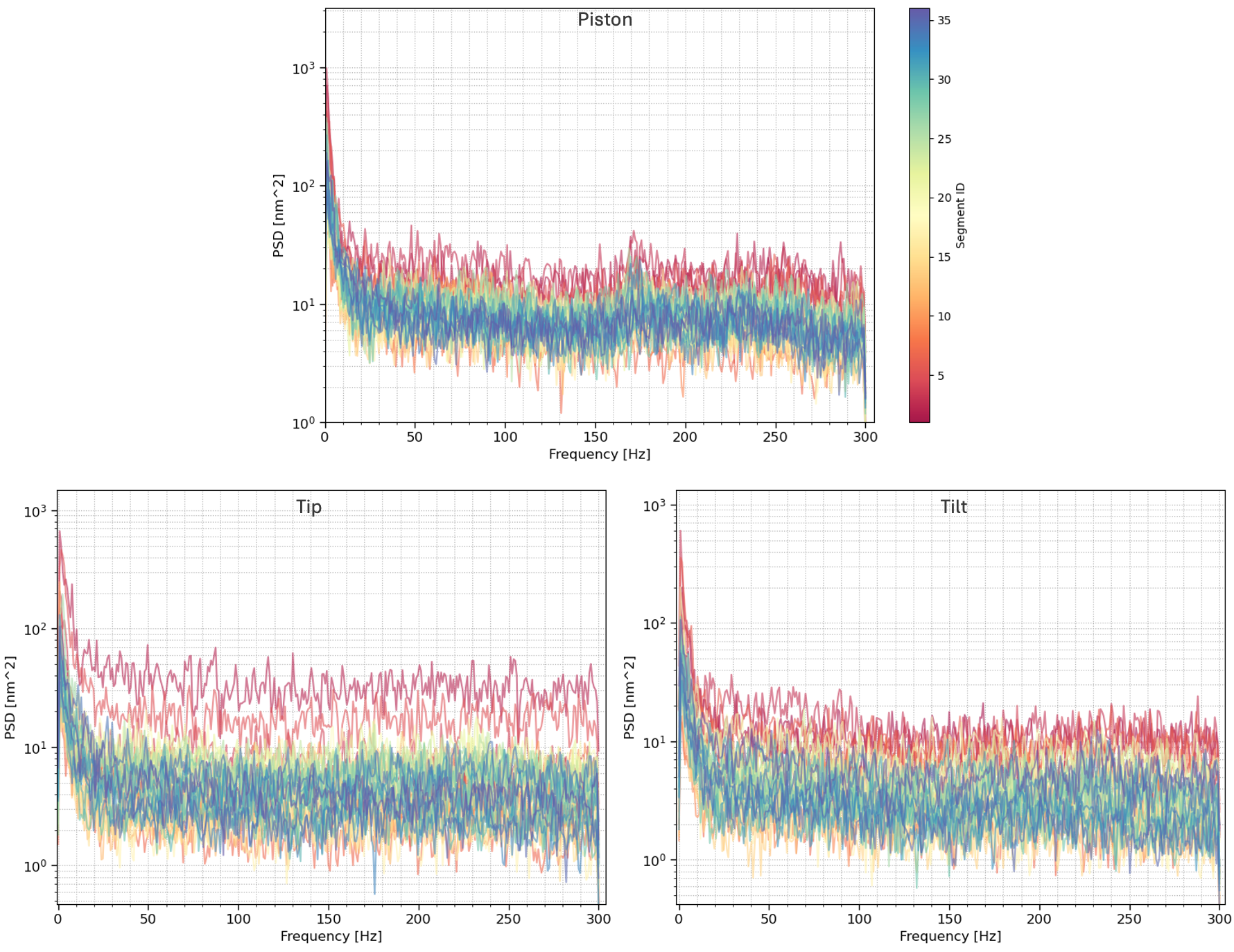}
    \caption{Power spectral density (PSD) distributions as a function of frequency for each segment piston, tip, and tilt measurements from the vZWFS signal taken at 600 Hz. The PSDs were estimated using Welch's method.}
    \label{fig:fast}
\end{figure}

The vZWFS currently installed at Keck can also be used for other diagnostics beyond segment phasing of the primary mirror. To this end, we ran a test pointing the telescope to a very bright star (Arcturus, Hmag = -2.8) and reading out the CRED2 camera at a speed of 600 Hz, much faster than the 2 Hz required for the segment-phasing closed-loop iterations. The vZWFS can then be used as a diagnostic to identify dynamics and vibrations in the system. Figure \ref{fig:fast} shows the power spectral density (PSD) distributions of piston, tip, and tilt measurements for each segment as a function of frequency. There may be a small piston peak at $\sim$175 Hz, though more data will be needed to confirm and further investigate its origin.

\section{Conclusion}
\label{sec:Concl} 

Using the vZWFS signal to control the Keck primary mirror segments, we have now performed a total of eight closed-loop runs, on five separate nights in 2023 August and 2024 February. They each resulted in a SR increase on the science PSF. These runs were conducted across a range of primary mirror conditions: during segment exchanges, immediately following telescope re-phasing, and weeks after the last re-phasing. For now, no clear pattern emerges, suggesting that the phasing solution reached by PCS with the edge sensors differs from the phasing solution reached by the post-AO vZWFS. More work is needed to disentangle the sources of aberrations, either originating from the primary mirror or rather being caused by AO effects. Further investigations of both questions will include: (1) determining optimal segment-phasing methodologies and (2) disentangling aberration sources in order to determine the optimal infrastructure, techniques, and control architectures needed to reach the high-contrast requirements for the next generation of planet hunting missions with the ELTs and space missions. 

The goal with the vZWFS on Keck is to improve the post-AO PSF quality and achievable contrasts on Keck in the short-term, while using Keck as a testbed to investigate questions relevant for the next generation of segmented large ground-based ELTs and space missions such as HWO. With the vZWFS as it is currently installed on Keck, we plan on conducting the following investigations: understanding what leads to the best AO performance for high-contrast imaging and determining the best ``phasing method", disentangling sources of aberrations, using the vZWFS to measure segment surface figures, correcting for segment tip and tilt in addition to piston, identifying any vibrations, and investigating the stability of the vZWFS closed-loop runs, both on the primary mirror segments and for the SHWFS slope offsets. In the long-term, the results of these tests will be translated into implications and recommendations for the ELTs and HWO. Although in different error regimes, the control architectures tested will be especially relevant to any system having a combination of edge sensors and post-AO wavefront measurements.

\acknowledgments % equivalent to \section*{ACKNOWLEDGMENTS}       

This work is funded by the Heising-Simons Foundation (grant \#2020-1822). Funding for KPIC has been provided by the California Institute of Technology, the Jet Propulsion Laboratory, the Heising-Simons Foundation (grants \#2015-129, \#2017-318, \#2019-1312, and \#2023-4598), the Simons Foundation (through the Caltech Center for Comparative Planetary Evolution), and the NSF under grant AST-1611623. Part of this research was carried out at the Jet Propulsion Laboratory, California Institute of Technology, under a contract with the National Aeronautics and Space Administration (80NM0018D0004).

The authors wish to recognize and acknowledge the very significant cultural role and reverence that the summit of Maunakea has always had within the indigenous Hawaiian community. We are most fortunate to have the opportunity to conduct observations from this mountain.

% References
\bibliography{report} % bibliography data in report.bib

\begin{thebibliography}{1}

\bibitem{Salama24}
{Salama}, M., {Guthery}, C., {Chambouleyron}, V., {Jensen-Clem}, R., {Wallace}, J.~K., {Delorme}, J.-R., {Troy}, M., {Wenger}, T., {Echeverri}, D., {Finnerty}, L., {Jovanovic}, N., {Liberman}, J., {L{\'o}pez}, R.~A., {Mawet}, D., {Morris}, E.~C., {van Kooten}, M., {Wang}, J.~J., {Wizinowich}, P., {Xin}, Y., and {Xuan}, J., ``{Keck Primary Mirror Closed-loop Segment Control Using a Vector-Zernike Wavefront Sensor},'' {\em The Astrophysical Journal (ApJ)}~{\bf 967},  171 (June 2024).

\bibitem{Mawet16}
{Mawet}, D., {Wizinowich}, P., {Dekany}, R., {Chun}, M., {Hall}, D., {Cetre}, S., {Guyon}, O., {Wallace}, J.~K., {Bowler}, B., {Liu}, M., {Ruane}, G., {Serabyn}, E., {Bartos}, R., {Wang}, J., {Vasisht}, G., {Fitzgerald}, M., {Skemer}, A., {Ireland}, M., {Fucik}, J., {Fortney}, J., {Crossfield}, I., {Hu}, R., and {Benneke}, B., ``{Keck Planet Imager and Characterizer: concept and phased implementation},'' in [{\em Adaptive Optics Systems V}{\nolinebreak\hspace{0.1em}]},  {Marchetti}, E., {Close}, L.~M., and {V{\'e}ran}, J.-P., eds., {\em Society of Photo-Optical Instrumentation Engineers (SPIE) Conference Series} {\bf 9909},  99090D (July 2016).

\bibitem{Delorme21}
{Delorme}, J.-R., {Jovanovic}, N., {Echeverri}, D., {Mawet}, D., {Kent Wallace}, J., {Bartos}, R.~D., {Cetre}, S., {Wizinowich}, P., {Ragland}, S., {Lilley}, S., {Wetherell}, E., {Doppmann}, G., {Wang}, J.~J., {Morris}, E.~C., {Ruffio}, J.-B., {Martin}, E.~C., {Fitzgerald}, M.~P., {Ruane}, G., {Schofield}, T., {Suominen}, N., {Calvin}, B., {Wang}, E., {Magnone}, K., {Johnson}, C., {Sohn}, J.~M., {L{\'o}pez}, R.~A., {Bond}, C.~Z., {Pezzato}, J., {Sayson}, J.~L., {Chun}, M., and {Skemer}, A.~J., ``{Keck Planet Imager and Characterizer: a dedicated single-mode fiber injection unit for high-resolution exoplanet spectroscopy},'' {\em Journal of Astronomical Telescopes, Instruments, and Systems}~{\bf 7},  035006 (July 2021).

\bibitem{Wizinowich13}
{Wizinowich}, P., ``{Astronomical Science with Adaptive Optics at the W. M. Keck Observatory},'' {\em Publications of the Astronomical Society of the Pacific (PASP)}~{\bf 125},  798 (July 2013).

\bibitem{Doelman19}
{Doelman}, D.~S., {Fagginger Auer}, F., {Escuti}, M.~J., and {Snik}, F., ``{Simultaneous phase and amplitude aberration sensing with a liquid-crystal vector-Zernike phase mask},'' {\em Optics Letters}~{\bf 44},  17 (Jan. 2019).

\bibitem{Wallace23}
{Wallace}, J., {Wenger}, T., {Jewell}, J., {Salama}, M., {Chambouleyron}, V., {van Kooten}, M., {Guthery}, C., {Ragland}, S., {Delorme}, J.-R., {Jensen-Clem}, R., {Wizinowich}, P., {Mawet}, D., and {Jovanovic}, N., ``{Architecting, Implementing and Observing with a Metasurface vector Zernike wavefront sensor on the Keck Telescope},'' in [{\em Seventh International Conference on Adaptive Optics for Extremely Large Telescopes (AO4ELT)}{\nolinebreak\hspace{0.1em}]},   9 (June 2023).

\bibitem{Chambouleyron24}
Chambouleyron, V., Salama, M., Cisse, M., Haffert, S.~Y., Deo, V., Guthery, C.~E., Wallace, J.~K., Dillon, D., Jensen-Clem, R., , Hinz, P., and Macintosh, B., ``{Evaluating various reconstruction techniques for the Zernike Wavefront Sensor: from simulations to on-sky tests},'' in [{\em In this proceeding}{\nolinebreak\hspace{0.1em}]},  {\em Society of Photo-Optical Instrumentation Engineers (SPIE) Conference Series} (July 2024).

\bibitem{Cohen94}
{Cohen}, R.~W., {Mast}, T.~S., and {Nelson}, J.~E., ``{Performance of the W.M. Keck telescope active mirror control system},'' in [{\em Advanced Technology Optical Telescopes V}{\nolinebreak\hspace{0.1em}]},  {Stepp}, L.~M., ed., {\em Society of Photo-Optical Instrumentation Engineers (SPIE) Conference Series} {\bf 2199},  105--116 (June 1994).

\bibitem{Chanan94}
{Chanan}, G.~A., {Nelson}, J.~E., {Mast}, T.~S., {Wizinowich}, P.~L., and {Schaefer}, B.~A., ``{W.M. Keck Telescope phasing camera system},'' in [{\em Instrumentation in Astronomy VIII}{\nolinebreak\hspace{0.1em}]},  {Crawford}, D.~L. and {Craine}, E.~R., eds., {\em Society of Photo-Optical Instrumentation Engineers (SPIE) Conference Series} {\bf 2198},  1139--1150 (June 1994).

\end{thebibliography}
\bibliographystyle{spiebib} % makes bibtex use spiebib.bst

\end{document}